\documentclass[twocolumn,prl,aps,epsfig,superscriptaddress,showpacs,amsmath,amssymb,floatfix]{revtex4}

\usepackage{epsfig}

\begin{document}

\title{Pseudogap and Fermi-arc Evolution in the Phase-fluctuation Scenario}

\author{Qiang Han}
\affiliation{Department of Physics, Renmin University of China, Beijing, China}
\affiliation{Department of Physics and Center of Theoretical and Computational Physics, The University of Hong Kong, Pokfulam Road, Hong Kong, China}

\author{Tao Li}
\affiliation{Department of Physics, Renmin University of China, Beijing, China}

\author{Z. D. Wang}
\affiliation{Department of Physics and Center of Theoretical and Computational Physics, The University of Hong Kong, Pokfulam Road, Hong Kong, China}

\date{\today}
\begin{abstract}
Pseudogap phenomena and the formation of Fermi arcs in underdoped cuprates are numerically studied in the presence of phase fluctuations that are simulated by an XY model. Most importantly the spectral function for each Monte Carlo sample is calculated directly and efficiently by the Chebyshev polynomials without having to diagonalize the fermion Hamiltonian, which enables us to handle a system large enough to achieve sufficient momentum/energy resolution.
We find that the momentum dependence of the energy gap is identical to that of a pure $d$-wave superconductor well below the KT-transition temperature ($T_\text{KT}$), while displays an upturn deviation from $\cos k_x - \cos k_y$ with increasing temperature. An abrupt onset of the Fermi arcs is observed above $T_\text{KT}$ and the arc length exhibits a similar temperature dependence to the thermally activated vortex excitations.

\end{abstract}

\maketitle

One of the most important open issues in the research of high temperature superconducitity is the origin of the so-called pseudogap and its relation to the superconducting gap. One viewpoint regards the pseudogap as preformed Cooper-pair state without long-range phase coherence, yet another attributes its origin to a new phase of matter. The most direct evidence of the pseudogap in hole-doped cuprates comes from experimental techniques such as angle resolved photoemission spectroscopy (ARPES)  and scanning tunneling spectroscopy (STS), which measure the low-energy single-particle spectrum and relate the loss of spectral weight to the pseudogap opening.~\cite{marshall,loeser96,ding96} One peculiar property of the pseudogap phase revealed by ARPES is the truncated Fermi surfaces termed as 'Fermi arcs'.~\cite{normannature98} The Fermi arcs emerge abruptly around the nodal region above $T_c$, and extend out to the Brillouin zone edge with increasing temperature and form a complete Fermi surface for temperature above $T^*$.\cite{normannature98,kanigel07,kanigel} This evolution of the Fermi surface shows a sharp distinction from that of conventional BCS superconductors.

The evolution of the Fermi surface in the pseudogap phase has been addressed \cite{berg,alvarez,tli} within the phase fluctuation context. The formation of Fermi arcs was attributed to the pile-up of low-energy spectral weight due to the Doppler effect of the fluctuating suppercurrent.~\cite{berg} However such analytic results are based on a semiclassical approximation \cite{franz} which was argued to be uncontrolled \cite{tsvelik}. Numerical treatment of the effect of phase fluctuations involved a combination of the Monte Carlo (MC) simulation of the Landau-Ginzburg free energy and the exact diagonalization of the fermion Hamiltonian.\cite{alvarez} However, the competition between the superconducting and antiferromagnetic order parameters in the presence of quenched disorder was complicated and the momentum resolution was relative low due to the time-consuming numerical diagonalization procedure.

In this work, we 1) examine the momentum and temperature dependence of the spectral gap and 2) temperature dependence of the Fermi-arc length within the phase-fluctuating scenario. The fluctuating phases are simulated phenomenologically by a classical XY model.~\cite{eckl,ychen} The $XY$ free energy is
\begin{equation}
F(\{\varphi_{\mathbf{i}}\})=-\frac{J}{2}\sum_{\mathbf{i},\boldsymbol{\delta}}\cos(\varphi_{\mathbf{i}}-\varphi_{\mathbf{i}+\boldsymbol{\delta}}),
\label{xymodel}
\end{equation}
where the phase-stiffness constant $J$ is the energy scale of the Kosterlitz-Thouless transition. $\varphi_\mathbf{i}$ denotes the phase of the superconducting order parameter defined on the lattice site $\mathbf{i}$. $\boldsymbol{\delta}$ is the unit vector along $x$ or $y$ direction connecting $\mathbf{i}$ to its nearest neighbors(NN).
The Hamiltonian of non-interacting electrons moving in a spatially fluctuating pairing field as described by Eq.~(\ref{xymodel}) reads
\begin{eqnarray}
\hat{H}&=&\sum_{\mathbf{i,j},\sigma} (-t_{\mathbf{ij}}-\mu\delta_{\mathbf{ij}}) c_{\mathbf{i}\sigma}^\dagger c_{\mathbf{j}\sigma}
     + \sum_{\mathbf{i,j}} (\Delta_{\mathbf{ij}} c_{\mathbf{i}\uparrow}^\dagger c_{\mathbf{j}\downarrow}^\dagger + H.c.), \nonumber \\
     &=& \sum_{\mathbf{i,j}}(c_{\mathbf{i}\uparrow}^\dagger, c_{\mathbf{i}\downarrow})
     \left( \begin{array}{cc}
     -t_{\mathbf{ij}}-\mu\delta_{\mathbf{ij}} & \Delta_{\mathbf{ij}}  \\
     \Delta^*_\mathbf{ij} & t_{\mathbf{ij}}+\mu\delta_{\mathbf{ij}}
     \end{array} \right)
     \left(
     \begin{array}{c}
        c_{\mathbf{j}\uparrow} \\
        c_{\mathbf{j}\downarrow}^\dagger
     \end{array}
     \right)
     \label{ham}
\end{eqnarray}
where $t_{\mathbf{ij}}$ denotes hopping integral. and $\mu$ represents the chemical potential. The pairing potential $\Delta_{\mathbf{ij}}$ is defined on the NN bond $(\mathbf{i},\mathbf{j=i}+\boldsymbol{\delta})$.
Here we fix the amplitude of the pairing potential while concentrate on the role played by the phase fluctuation, i.e $\Delta_{\mathbf{i},\mathbf{i}+\boldsymbol{\delta}}=\Delta e^{i\theta_{\boldsymbol{\delta}}(\mathbf{i})}$ with $\Delta$ kept constant. We use the expression  $\theta_{\hat{x},\hat{y}}(\mathbf{i})=\pm \text{Im}\ln(e^{i\varphi_\mathbf{i}}+e^{i\varphi_{\mathbf{i}+\boldsymbol{\delta}}})$ to relate the phase of the pairing potential with $\varphi_{\mathbf{i}}$, which gives rise to pure $d$-wave pairing state in the low temperature limit when the thermal fluctuation is negligible small.

In our numerical study, a set of phase configurations $\{ \varphi_{\mathbf{i}} \}$ is first generated by the classical Monte Carlo simulation according to the probability distribution $\exp(-F(\{\varphi_\mathbf{i}\})/T$). Then the single-electron spectral function of the BdG Hamiltonian (\ref{ham}) for each individual phase configuration, which is denoted by $A_\varphi(\mathbf{k},\omega)$ in short, is calculated and at last averaged over the sampled phase space to obtain the expectation value of the spectral function  $A(\mathbf{k},\omega)=\langle A_\varphi(\mathbf{k},\omega)\rangle$. The bottleneck in this algorithm mainly lies in the numerical calculation of the spectrum of the BdG Hamiltonian (\ref{ham}). Previous studies \cite{eckl,alvarez} attempted to solve the problem by direct diagonalization of the BdG Hamiltonian, whose workload is $O(N^3)$ with $N=L^2$ the lattice size and rather time-consuming when $L$ is large. In this paper, we deploy the Chebyshev expansion approach \cite{Weibe} to directly calculate the spectral function, which circumvents the difficulty in fully diagonalizing a large Hermitian matrix. As illustrated in the following, the computational load for each $A_\varphi(\mathbf{k},\omega)$ is approximately $O(MN)$ with $M<<N$, which is a significant advance regarding to numerical studies of the above problem.

The single-particle spectral function corresponding to the quadratic Hamiltonian of Eq.~(\ref{ham}) can be calculated from the imaginary part of the Green's function
\begin{equation}
A_\varphi(\nu,\omega)=-\frac{1}{\pi}\mbox{Im}\langle\nu| (\omega+i\eta-\hat{H}_\varphi)^{-1} |\nu\rangle.
\end{equation}
Here more generally, $|\nu\rangle$ denotes any single-particle wave vector concerned with. $|\nu\rangle$ can be chosen as $|\mathbf{k}\rangle=\frac{1}{\sqrt{N}}\sum_{\mathbf{i}}e^{i\mathbf{k}\cdot\mathbf{i}}|\mathbf{i}\rangle$ with $\mathbf{k}$ the momentum in the first Brillouin zone, and accordingly the obtained $A(\mathbf{k},\omega)$ are related to the evolution of the Fermi surface as observed by ARPES. If
$|\nu\rangle = |\mathbf{i}\rangle$, $A(\mathbf{i},\omega)$ is actually the local density of state related to the STS measurements. According to the kernel polynomial method \cite{Weibe},
$A_\varphi(\nu,\omega)$ can be expanded in truncated series of the Chebyshev polynomials of the first kind $T_m$,
\begin{equation}
A_\varphi(\nu,\omega)=-\frac{\mu_0+2\sum_{m=1}^{M-1}\mu_m g_m T_m(\omega/s)}{\pi\sqrt{1-(\omega/s)^2}},
\end{equation}
where the expansion coefficients
\begin{equation}
    \mu_m=\langle\nu|T_m(\hat{H}/s)|\nu\rangle,
    \label{moment}
\end{equation}
are Chebyshev moments. $g_m=\sinh[\lambda(1-m/M)]/\sinh(\lambda)$ is
the Lorentz kernel factor which improves the convergence of the truncated series and damps Gibbs oscillations. A good value of the kernel parameter is $\lambda=4$. The scaling factor $s$ fits the spectrum of the Hamiltonian $\hat{H}/s$ into the interval $[-1,1]$. We choose $s$ to be a slightly larger than the bandwidth of the BdG Hamiltonian (\ref{ham}). The energy resolution of the spectral function is estimated by $\varepsilon=\lambda s/M$. Most computational effort is spent in the calculation of the moments $\mu_m$ according to Eq.~(\ref{moment}), which reduces to matrix-vector multiplications after taking advantage of the recursion relation $T_m(x)=2xT_{m-1}(x)-T_{m-2}(x)$. Considering that the BdG Hamiltonian (\ref{ham}) is sparse, the cost of matrix-vector multiplication is an order $O(N)$ process and the calculation of $M$ moments requires only $O(MN)$ computational operations, which even enables us to handle larger lattices using a desktop computer. Furthermore, recursive relations of Chebyshev polynomials $T_{2m}=2T_m^2-1$ and $T_{2m+1}=2T_mT_{m+1}-T_1$ enable us to obtain two moments per matrix-vector multiplication.

\begin{figure}[ht]
\begin{tabular}{cc}
\epsfig{figure=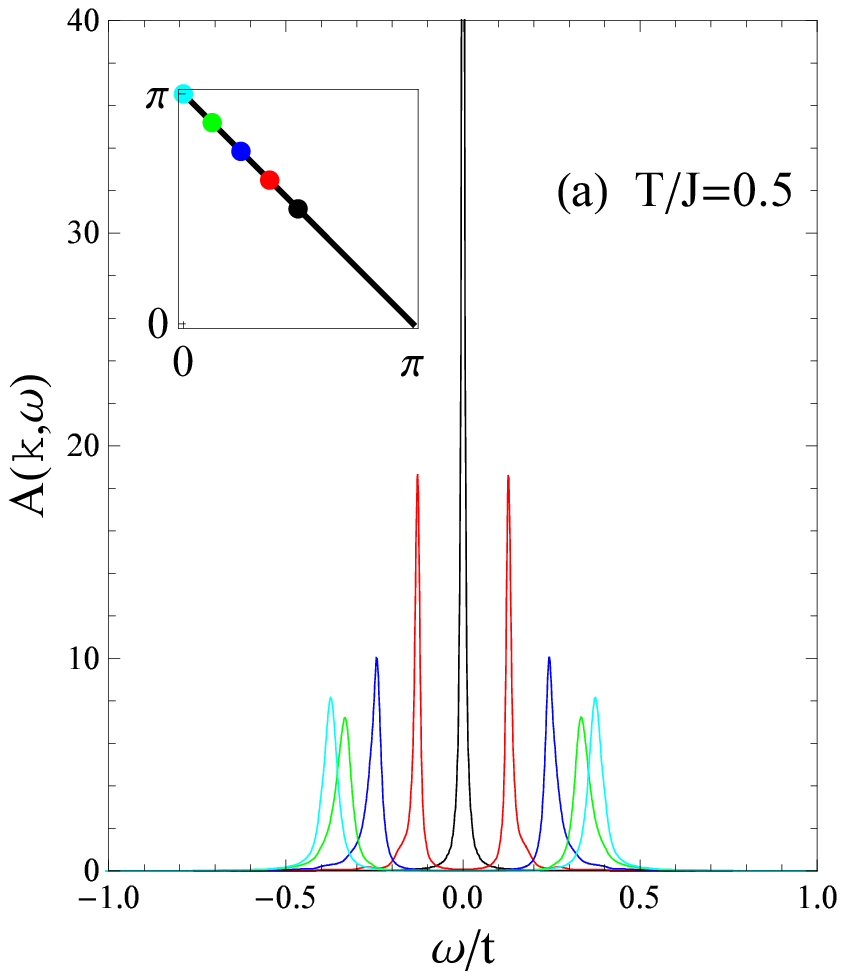,width=0.22\textwidth} & \epsfig{figure=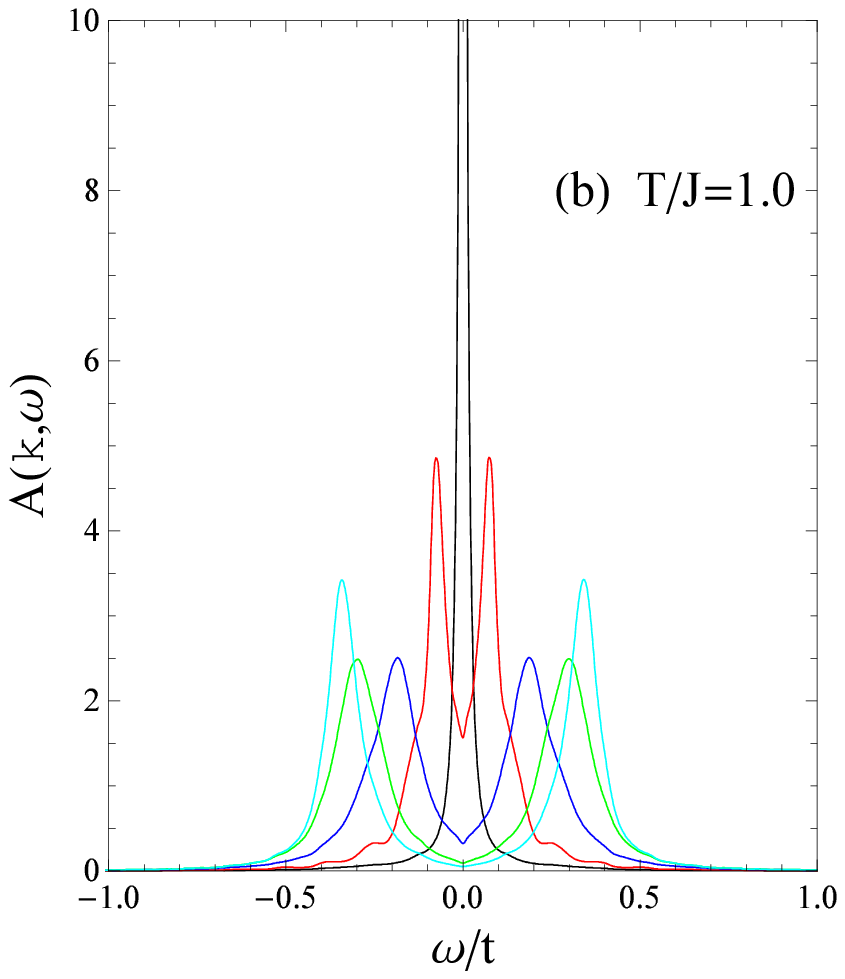,width=0.22\textwidth} \\ \epsfig{figure=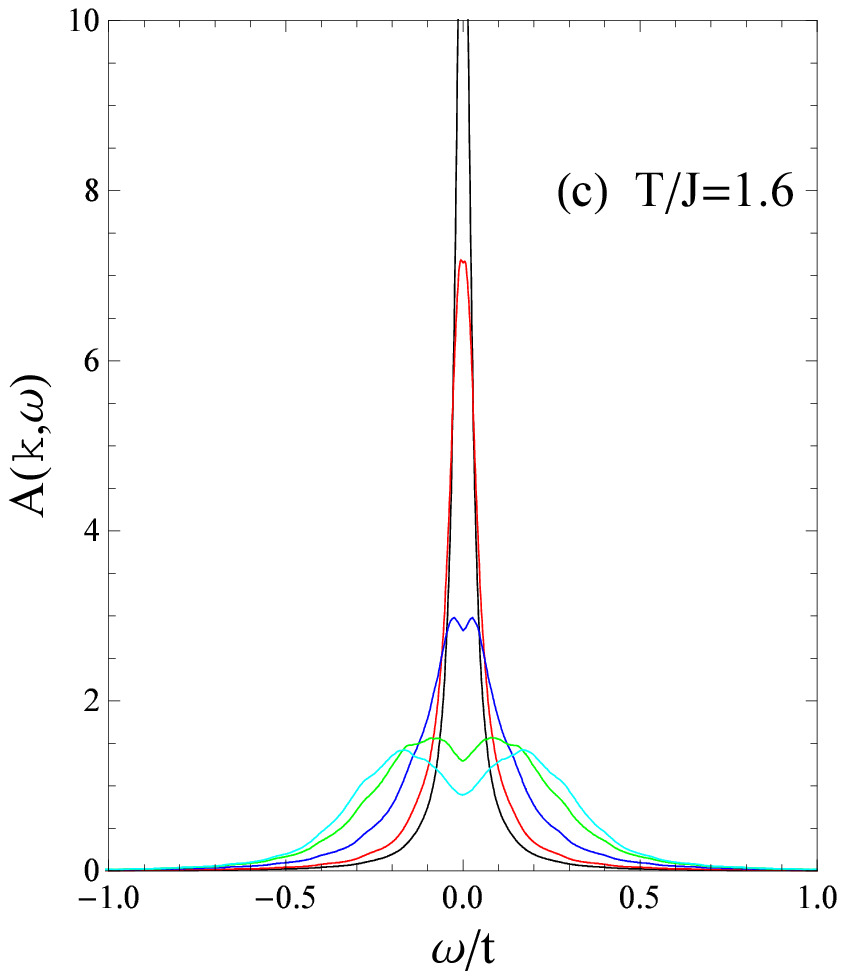,width=0.22\textwidth} & \epsfig{figure=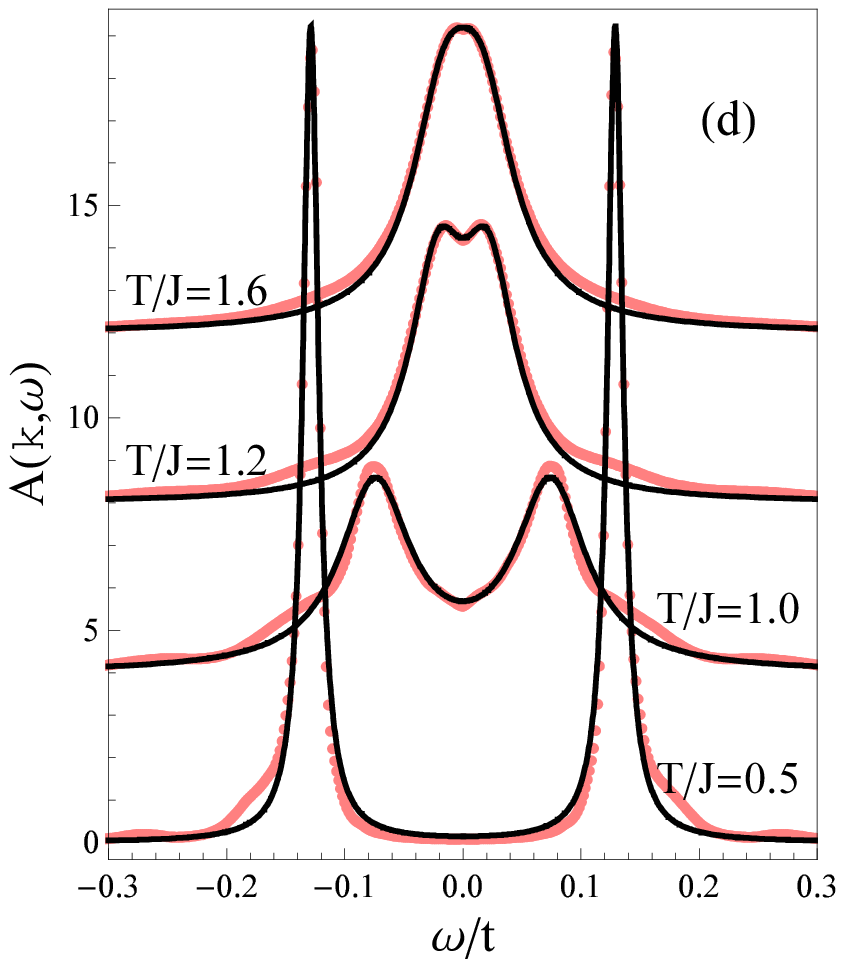,width=0.22\textwidth}
\end{tabular}
\caption{Energy and momentum dependence of the spectral function (a), at $T/J=0.5$ in the superconducting state, (b), at $T/J=1.0$ around the KT phase transition and (c), at $T/J=1.6$ in the pseudogap state. The inset of (a) shows the Fermi surface (black line) in first quarter of the first Brillouin zone together with some selected Fermi momenta whose colors correspond to those of the curves in (a)-(c). (d), Variation of the spectral function with temperature for fixed Fermi momentum (red dot in the inset) superimposed by their fits (black curves) according to Eq.~(\ref{selfenergy}) }
\label{ATk}
\end{figure}

We now present the results of our simulations. Typical calculation is done on a $96\times 96$ lattice. Combined Monte Carlo method is used to generate equilibrium configurations of phase angles $\{\varphi_i\}$. The KT-phase transition is observed with $T_\text{KT}\approx J$ by studying the phase correlation and the vortex density as a function of temperature, which is consistent with our previous work.~\cite{tli} For each temperature, the first $10^5$ MC sweeps are dropped to equilibrate the system. $1000$ phase configurations every $1000$ MC sweeps are used to get thermal averages, which are sufficient to reduce the statistical errors. For each phase configuration, the corresponding spectral function is calculated by Chebyshev expansions with the truncation $M=4096$, which results in an approximate energy resolution $\varepsilon/t\simeq 0.004$. Here $t$ denotes the NN hopping integral. Most calculations are performed for the particle-hole symmetric case ($t^\prime=\mu=0$) for simplicity. $\Delta=0.1t$ which results in the superconducting gap $0.4t$ for the particle-hole symmetric case in the zero temperature limit. The resulting Fermi surface has a diamond shape, part of which is shown in the inset of Fig.~\ref{ATk}(a). On the node-to-antinode segment of the Fermi surface, 49 equally-spaced Fermi momenta are chosen to achieve sufficient momentum resolution.

Figure \ref{ATk} shows the energy distribution of the spectral function $A(\mathbf{k},\omega)$ for several Fermi momenta for illustration.
Well below $T_\text{KT}$ with $T/J=0.5$ we find sharp spectral peaks, indicating well-defined Bogoliubov quasiparticle excitations. Furthermore the superconducting gap vanishes only at the node, away from which the gap gradually opens and increases to its full size at the antinodal point. With increasing temperature from Fig.~\ref{ATk}(a)-(c), the spectral lines are broadened due to the strengthened scattering effect from the pair fluctuations. The energy gaps of $\mathbf{k}$'s near the antinode (see the evolution of the green and cyan curves) keep opening while that near the node (see red curves of (a)-(c) and (d)) closes as temperature rises across $T_\text{KT}\approx J$, indicating the formation of pseudogap at the antinodal region and the growth of Fermi arc near the nodal region above $T_\text{KT}$.
From the calculated spectral function in momentum-energy space, one can suggest two types of gap: the spectral gap $\Delta_{sg}(\mathbf{k})$ which is half of the peak-peak separation \cite{damascelli} and the fitted gap $\Delta_{fg}(\mathbf{k})$ by numerical fitting with a simple phenomenological model
$A(\mathbf{k},\omega)=-\mbox{Im}[\omega-\xi_\mathbf{k}-\Sigma(\mathbf{k},\omega)]^{-1}/\pi$ with the modified BCS self-energy,\cite{schrieffer,norman98,norman07}
\begin{equation}
\Sigma(\mathbf{k},\omega)=-i\Gamma_\mathbf{k}+
    \frac{\Delta^2_{fg}(\mathbf{k})}{\omega+\xi_\mathbf{k}+i\Gamma_\mathbf{k}},
    \label{selfenergy}
\end{equation}
where $\xi_\mathbf{k}=-2t(\cos k_x + \cos k_y)-4t^\prime\cos k_x \cos k_y - \mu$ the normal-state electron dispersion relation. $\Delta_{fg}(\mathbf{k})$ and $\Gamma_\mathbf{k}$ are fitting parameters. $\Gamma_\mathbf{k}$ takes into account of the lifetime broadening due to the scattering effect from the pair fluctuations. Our model of a momentum-dependent scattering rate $\Gamma_\mathbf{k}$ is more general than that considered in Ref.~\cite{norman98,norman07}. Fig.~\ref{ATk}(d) shows the typical fitting results together with the raw data calculated from Monte Carlo simulations.

\begin{figure}
\begin{tabular}{cc}
    \epsfig{figure=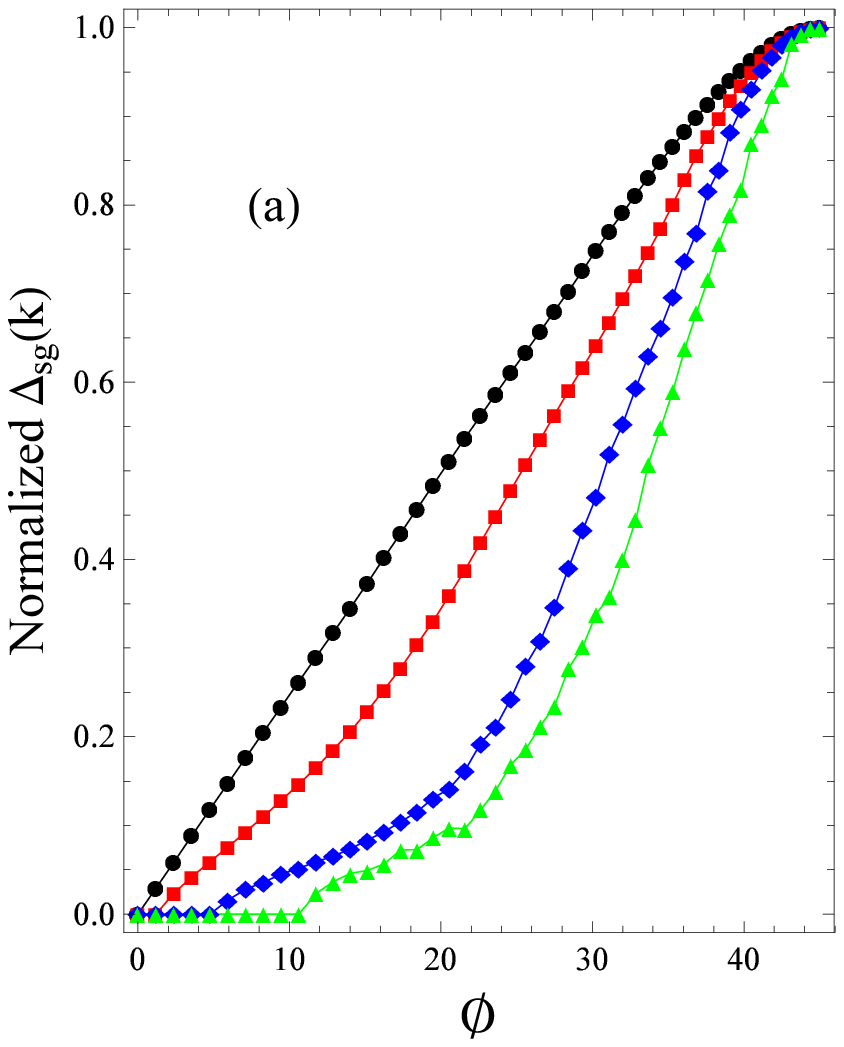,width=0.22\textwidth} & \epsfig{figure=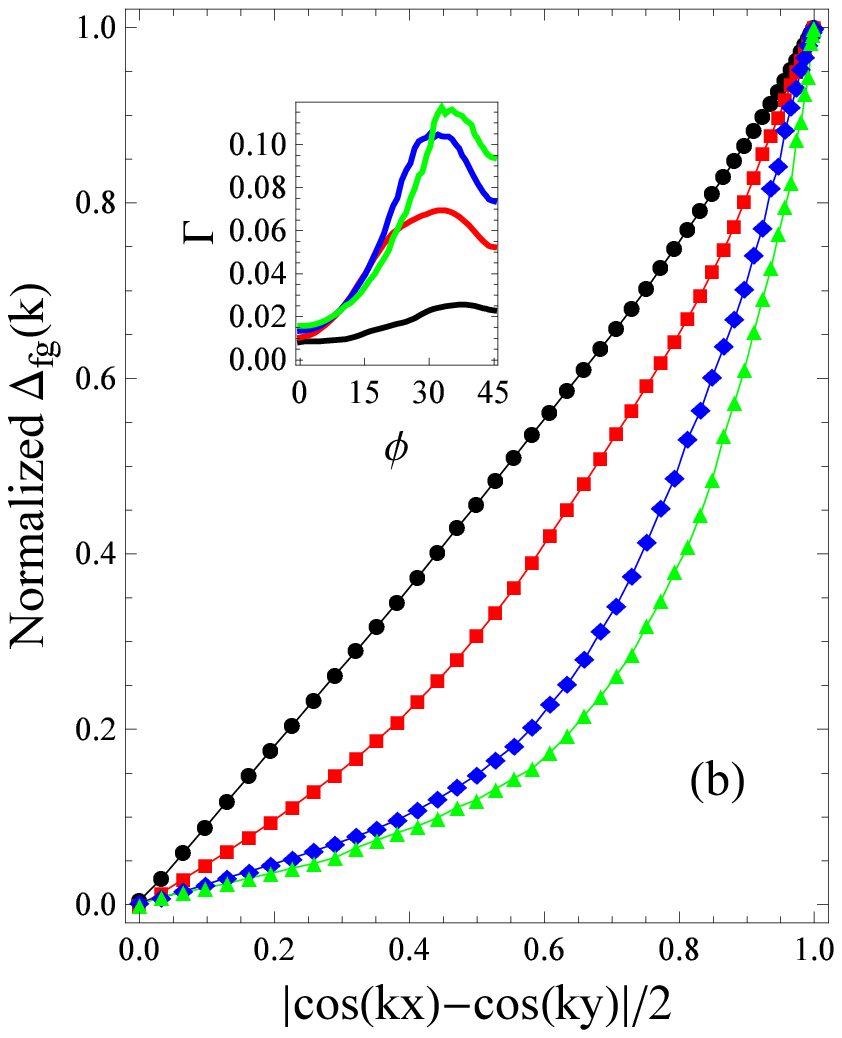,width=0.22\textwidth}
    \end{tabular}
\caption{
Momentum dependence of energy gap at four different temperatures $T/J=0.5, 1.0, 1.1, 1.2$(black dot, red square, blue diamond, and green triangle).
(a) spectral gap (half the peak-to-peak distance) plotted as a function of $\phi$, i.e.~the angle of $\mathbf{k}$ relative to the nodal point. (b) fitted gap (see Eq.~\ref{selfenergy}) plotted as a function of $|\cos(k_x)-\cos(k_y)|/2$. The gap is normalized to its value at the antinode. Inset of (b) shows the angle dependence of $\Gamma_\mathbf{k}$ (see Eq.~\ref{selfenergy})}
\label{gapk}
\end{figure}
Figure \ref{gapk} shows the dispersion of the spectral gap $\Delta_{sg}(\mathbf{k})$ and fitted gap $\Delta_{fg}(\mathbf{k})$. At $T/J=0.5$ (and below) both gap functions show simple $d_{x^2-y^2}$ form, $|\cos k_x - \cos k_y|/2$, indicating that the effect of the pair fluctuation is rather weak well below $T_\text{KT}$. With increasing temperature until $T/J=1.0$ which is approximately equal to $T_\text{KT}$, both gap functions deviate from the pure $d$-wave form and exhibit an upturn behavior due to increased spin-wave-type superconducting fluctuations.
This deviation increases more rapidly with temperature from $T/J=1.0$ to $1.2$, which indicates that the vortex-like excitations are more efficient in Doppler-shifting the single-electron spectral functions. From Fig.~\ref{gapk}(b), the behavior of $\Delta_{fg}$ changes gradually with temperature.
However by examining $\Delta_{sg}$ as in (a), we find that an emergent "arc", i.e. a Fermi segment with zero spectral gap, arises abruptly around $T_\text{KT}$ and grows rapidly with temperature from $T/J=1.0$ to $1.2$. This expansion of a Fermi node (below $T_\text{KT}$) to an Fermi arc (above $T_\text{KT}$) can be attributed to thermal proliferation of vortices.

\begin{figure}
\epsfig{figure=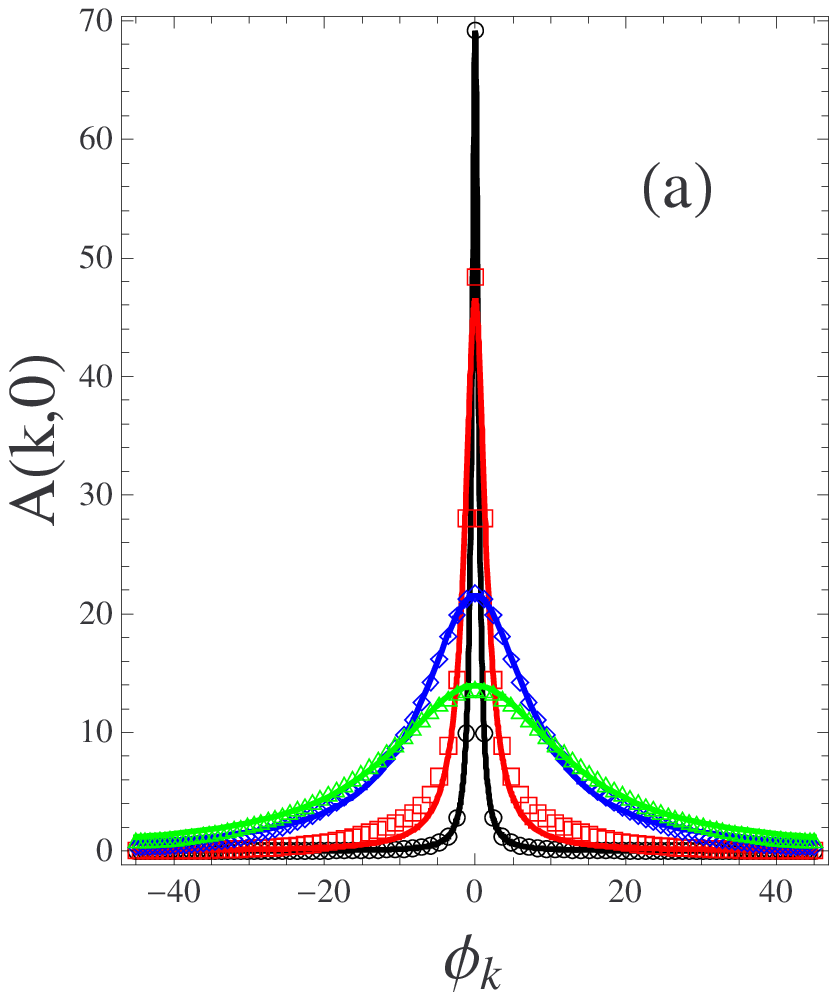,width=0.22\textwidth}
\epsfig{figure=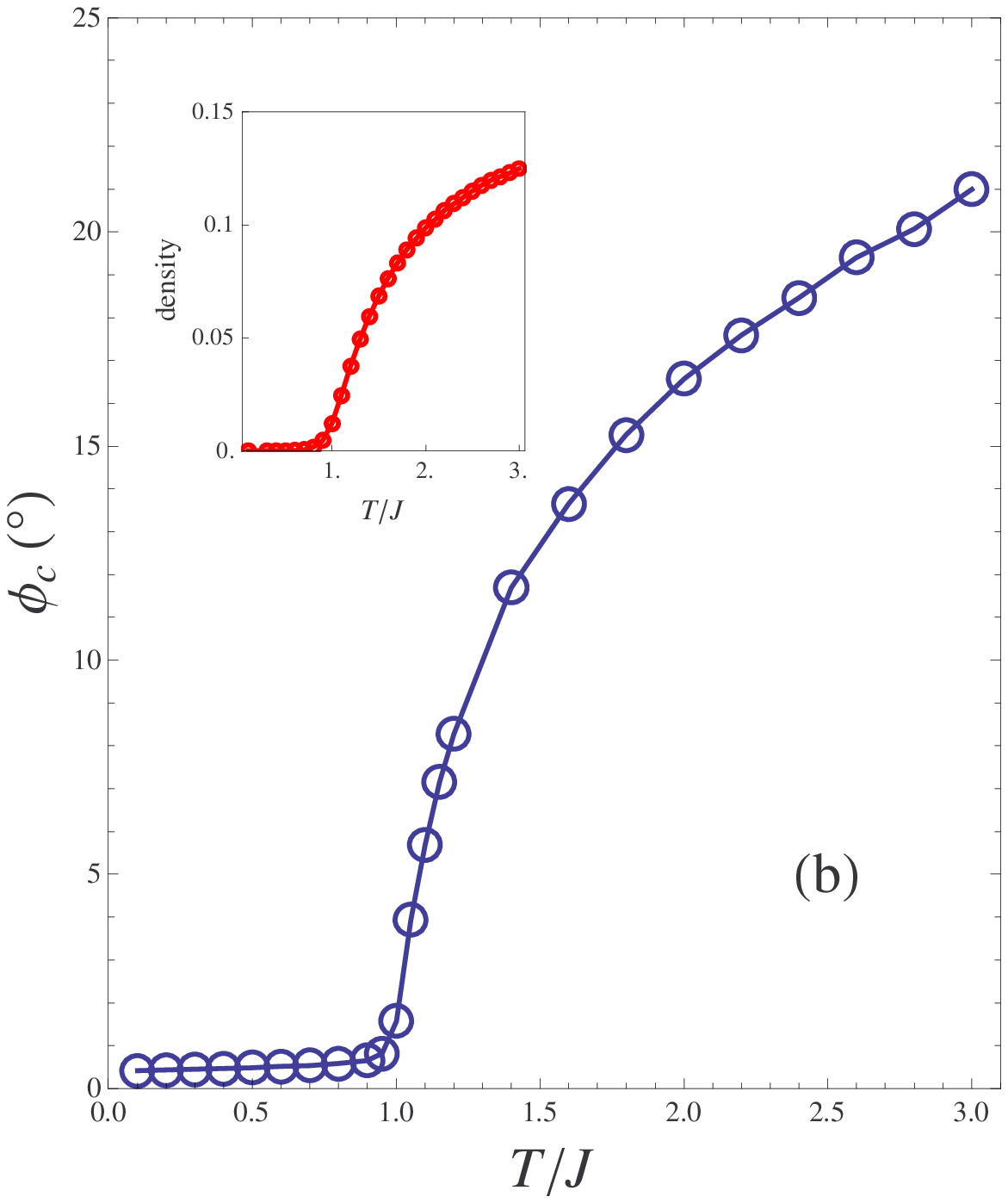,width=0.22\textwidth}
\caption{
(a), Momentum dependence of the spectral function at the Fermi level $A(\mathbf{k},\omega=0)$ with $\mathbf{k}$ along the underlying Fermi surface at four temperatures $T/J=0.5, 1.0, 1.2, 1.6$ (black circle, red square, blue diamond, green triangle). $\phi_\mathbf{k}$ denotes the angle of $\mathbf{k}$ relative to the node point $(\pi/2,\pi/2)$. Solid colored lines are fits to a Lorentzian-type function $\frac{a}{\phi_\mathbf{k}^2+\phi_c^2}$ (b),  the characteristic width $\phi_c$, as a function of temperature from $T/J=0.1$ to $3.0$. The inset shows the temperature dependence of vortex density.}
\label{arclength}
\end{figure}

The length of Fermi arc can be quantified by different ways. Experimentally one can locate the specific $\mathbf{k}$ where the spectral gap begins to open by examining the $\mathbf{k}$ dependence of $\Delta_{sg}(\mathbf{k})$, or by examining the loss of spectral weight.\cite{kanigel} Here we focus on the fact that Fermi arcs are segments of Fermi surface where the single-electron spectral function at the Fermi energy has non-vanishing value. Therefore $A(\mathbf{k},\omega=0)$ is studied with $\mathbf{k}$ along the underlying Fermi surface of the non-interacting electrons. The results are shown in Fig.~\ref{arclength}.
At $T/J=0.5$ we find a sharp peak around the nodal point, which decays rapidly along the node-to-antinode direction indicating point-like Fermi surface in the superconducting state. This peak is broadened with increasing temperature and accordingly phase fluctuations. Compared with the Gaussian-type decay obtained in Ref.~\cite{berg},  we find that $A(\mathbf{k},0)$ can be better fitted by a Lorentzian-type function $\frac{a}{\phi_\mathbf{k}^2+\phi_c^2}$ as shown in Fig.~\ref{arclength}(a), where $\phi_\mathbf{k}$ is the angle of $\mathbf{k}$ relative to the node, and $a$, $\phi_c$ are fitting parameters. This is not accidental but compatible with Eq.~(\ref{selfenergy}). For $\mathbf{k}$ along the Fermi surface and near the gap node, $\Delta_{fg}(\mathbf{k})\approx  (\sqrt{2}\pi-k_F)v_\Delta\phi_\mathbf{k}$ and after considering that $\xi_\mathbf{k}=0$ we have according to Eq.~(\ref{selfenergy}),
\begin{equation}
A(\mathbf{k},0)=\frac{\Gamma_\mathbf{k}/\pi}{\Delta_{fg}^2(\mathbf{k})+\Gamma_\mathbf{k}^2}\approx\frac{a}{\phi_\mathbf{k}^2+\phi_c^2}.
\end{equation}
Here $\phi_c$ can be regarded as the characteristic width of the decay of the spectral function and accordingly as a measure of the length of Fermi arc. Fig.~\ref{arclength}(b) shows the temperature dependence of $\phi_c$. Below $T_\text{KT}$ the arc length is negligible small, showing that the Fermi node is protected against weak phase fluctuations. Near $T/J=1.0$ there is an apparent jump of $\phi_c$, indicating that the Fermi arc emerges immediately above $T_\text{KT}$ together with the loss of phase coherence, which is consistent with experimental observation.~\cite{kanigel07} The relation of Fermi arc formation to the vortex proliferation is illustrated by the inset of Fig.~\ref{arclength}(b) where the temperature dependence of the vortex density is shown. The formation of vortices increases rapidly for temperatures near and above $T_\text{KT}$ and exhibits a similar temperature behavior to the arc length indicating that vortex-type phase fluctuations are responsible for the formation and evolution of the Fermi surface. This result is qualitatively consistent with an earlier theoretical study \cite{berg} in which the arc length is related to the characteristic width of the supercurrent distribution. However, our work has taken both the Doppler effect of the whirling supercurrent and the scattering effect of vortex centers into full consideration.

\begin{figure}[htb]
\epsfig{figure=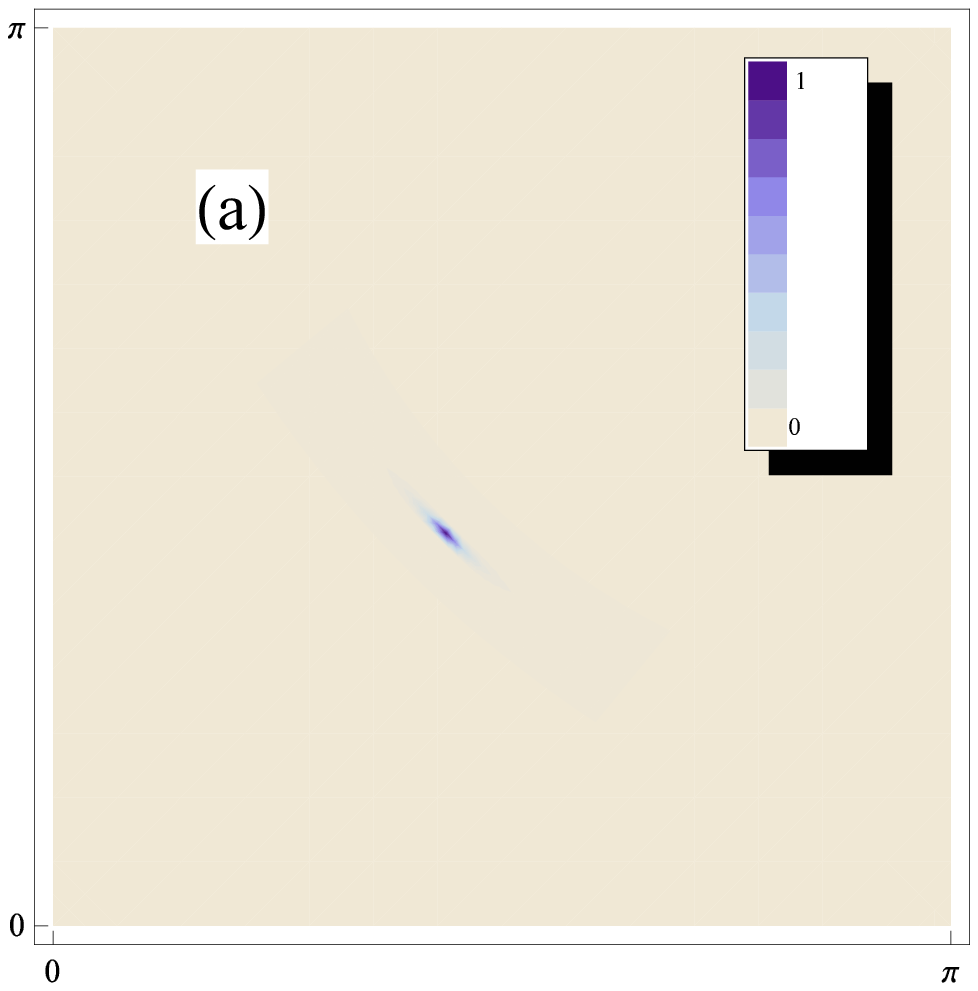,width=0.2\textwidth}
\epsfig{figure=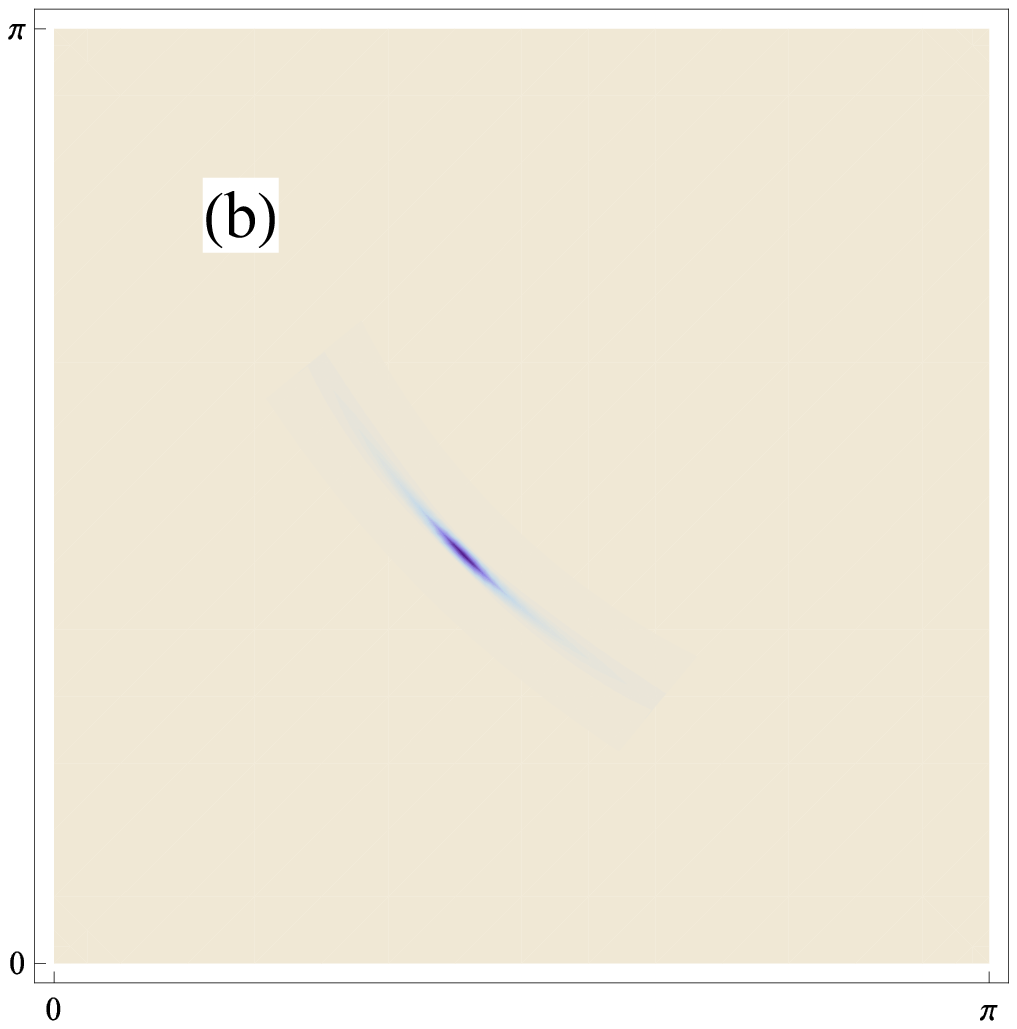,width=0.2\textwidth}
\epsfig{figure=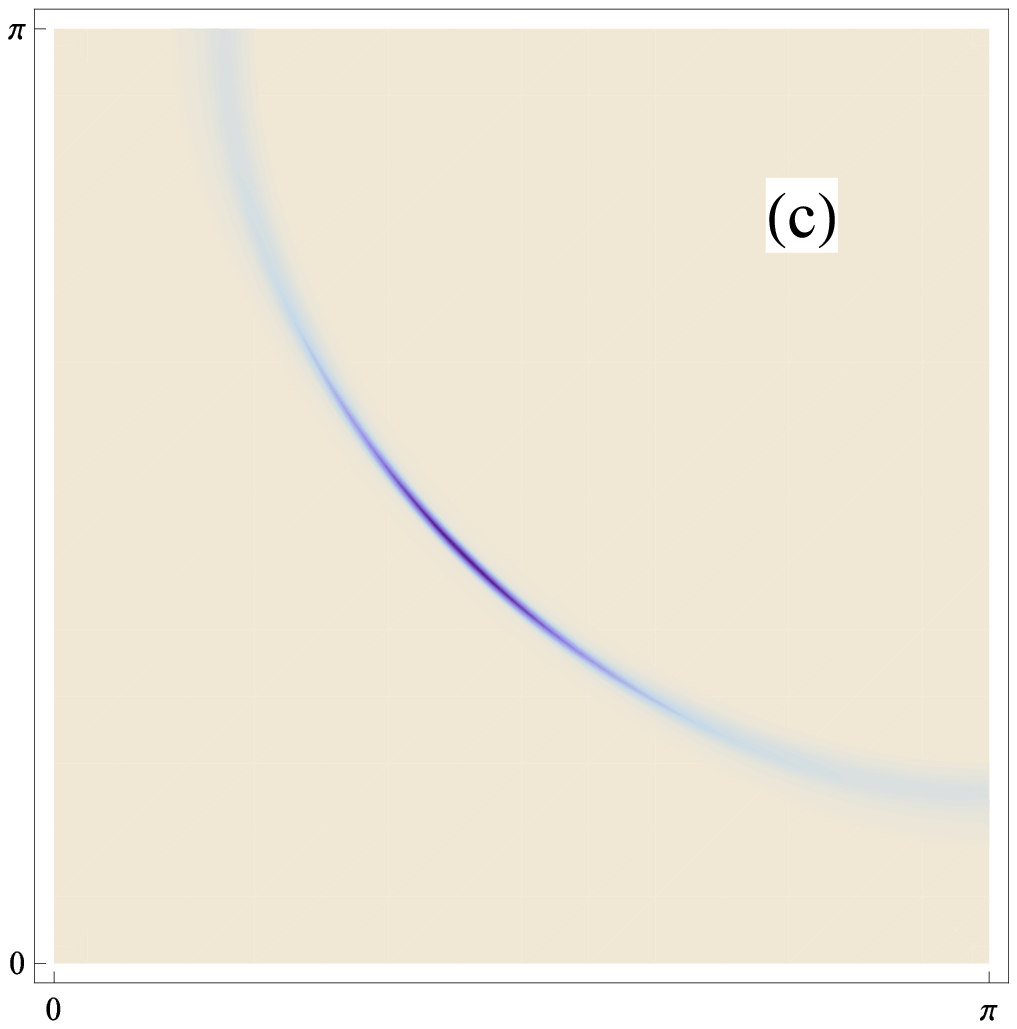,width=0.2\textwidth}
\epsfig{figure=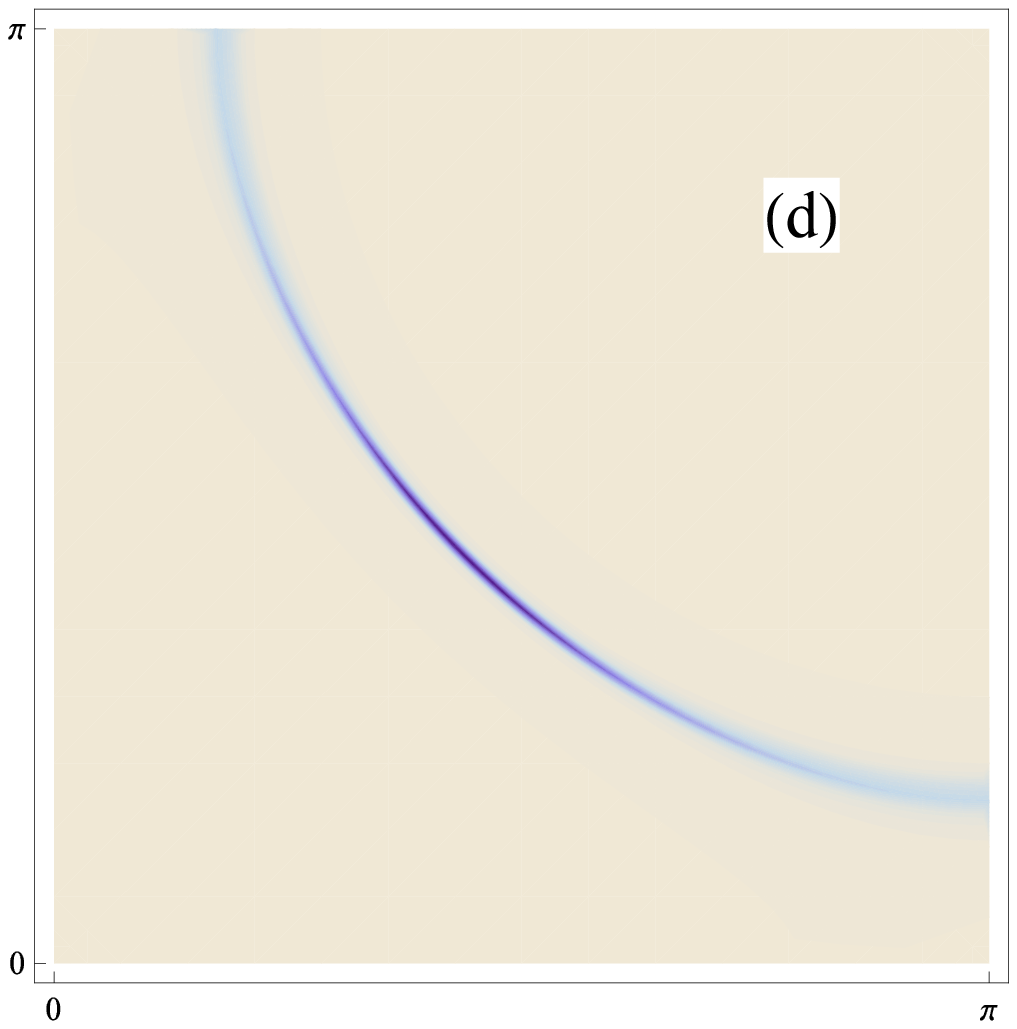,width=0.2\textwidth}
\caption{
The temperature dependence of the spectral function at the Fermi level, $A(\mathbf{k},\omega=0)$ normalized to its value at the node point for realistic band parameters $t^\prime/t=-0.3$. Here we only show the results in the first quarter of the first Brillouin zone. From (a) to (d), $T/J=0.5, 1.0, 1.6, 2.0$. }
\label{contour}
\end{figure}
In Fig.~\ref{contour} we plot the intensity of the spectral function at zero energy as a function of $k_x$ and $k_y$ in the first quarter of the first Brillouin zone for a realistic hopping parameter with $t^\prime/t=-0.3$. Well below the KT-transition temperature with ignorable phase fluctuations, the Fermi surface is point like as in ideal $d$-wave superconductors. With increasing temperature we find arc-like Fermi surface with monotonically increasing arc length, which is qualitatively consistent with ARPES observation \cite{kanigel07,kanigel}.

In summary, the influence of thermal phase fluctuation on the single-particle spectral function has been studied via numerical techniques which combine the Monte Carlo method with the Chebyshev expansion approach. The dispersion of the spectral gap deviates from the simple $d$-wave form especially for $T>T_{KT}$ when vortex-like fluctuations are dominant. Apparent onset of Fermi arcs are found near and above $T_{KT}$, with the arc length rises monotonically with temperature and displays close relevance to the vortex density. Our results give qualitative understanding of the ARPES observations of the pseudogap phase and Fermi arcs in cuprates. Quantitatively, specific temperature dependence of the phase stiffness $J$ is also crucial and its determination may calls for starting from a microscopic Hamiltonian. This issue will be addressed in the future work.

The work was supported by the RGC of Hong Kong under Grant No.HKU7055/09P, the URC fund of HKU, the Natural Science
Foundation of China No.10674179.

\bibliography{ref}

\end{document}